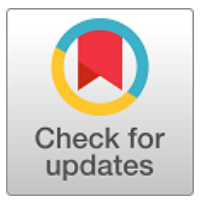

# Scalable low-latency entanglement distribution for distributed quantum computing

JIAPENG ZHAO,[1,*] YANG XU,[2] XIYUAN LU,[3] ENEET KAUR,[1] MICHAEL KILZER,[1] RAMANA KOMPELLA,[1] ROBERT W. BOYD,[2,4,5] AND REZA NEJABATI[1]

[1]Cisco Quantum Labs, 3232 Nebraska Ave, Santa Monica, California, 90404, USA
[2]Department of Physics and Astronomy, University of Rochester, Rochester, New York, 14627, USA
[3]Joint Quantum Institute, University of Maryland/NIST, College Park, Maryland, 20742, USA
[4]The Institute of Optics, University of Rochester, Rochester, New York, 14627, USA
[5]Department of Physics, University of Ottawa, Ottawa, Ontario, K1N 6N5, Canada
*penzhao2@cisco.com

**Practical distributed quantum computing and error correction require quantum networks with high-qubit-rate, high-fidelity, and low-reconfiguration-latency. Unfortunately, current approaches are limited by fundamental constraints: single-channel entanglement rates remain at the MHz level with millisecond-level reconfiguration, which is insufficient for fault-tolerant distributed quantum computing. Here, we propose a quantum network architecture that leverages reconfigurable quantum interfaces and wavelength-selective switches to overcome bandwidth and latency constraints. By tuning the frequency and temporal modes of photonic qubits across dense wavelength division multiplexing (DWDM) channels, our protocol achieves an entanglement generation rate of up to 183.4 MHz based on our comprehensive modeling of the networked cold atom computing systems. Our architecture enables nanosecond-scale network reconfiguration with low loss, low infidelity, and high dimensionality. Our modeling and simulation are designed for deployable distributed quantum computing and error correction, integrating the quantum interface, network switching, circuit compilation, and execution into a unified framework. The proposed architecture is fully compatible with industry-standard DWDM infrastructure, providing a scalable and cost-effective foundation for distributed quantum computing.**

## 1. INTRODUCTION

Quantum computers with millions of qubits are necessary to achieve practical fault-tolerant quantum computing. Some applications, such as factoring a 2,048-bit number using Shor's algorithm, are estimated to need around a million qubits [1]. However, practical hardware limitations make it extremely difficult to scale a single quantum processing unit (QPU) beyond $10^4$ qubits [2].

A promising approach to large-scale quantum computing is to interconnect multiple QPUs in a modular architecture. In such a quantum data center model, individual QPUs are linked by quantum interfaces, switches, and optical communication links to operate as a unified system [3,4]. The success of such a distributed quantum computing architecture is highly dependent on the high-rate and high-fidelity Bell pair distribution between arbitrary nodes, which enables gate teleportation, qubit teleportation, and distributed error correction to execute remote quantum circuits [5,6].

Unfortunately, most high-rate entanglement distribution architectures proposed previously rely on multiplexing broadband entanglement photons generated by spontaneous parametric down-conversion (SPDC) or spontaneous four-wave mixing (SFWM) [7–12]. Even though these entanglement sources naturally enable wavelength division multiplexing to boost the photon rate, they are not ideal for directly distributing entanglement between QPUs or memories. The first fundamental challenge comes from the stochastic nature of photons, which results in a low success probability of Bell state measurement (BSM) [3,10,12]. The second limitation is that the entanglement distribution between two QPUs requires two more BSMs when two entanglement sources are involved [3]. The third fundamental limitation comes from the temporal mode mismatch between broadband photons from an entanglement source, usually at GHz level or above, and computation qubits, typically MHz level. Therefore, entanglement photons from broadband SPDC or SFWM sources are not ideal for distributed quantum computing.

For comparison, entangled photons emitted directly from computation qubits are naturally optimal for distributing entanglement between QPUs, and experimental demonstra-tions have been realized in different quantum computing or memory modalities [4,13–16]. However, moving from these demonstrations of concept experiments to enable distributed quantum computing advantage still requires high-fidelity, low-loss, low-noise photonic links with high information capacity, embedded in a reconfigurable and dynamic network. Several major challenges currently hinder this approach:

1. Most computation qubits cannot be directly entangled with photonic qubits in the low-loss telecommunication band.
2. The temporal mode mismatch between photonic qubits and the transition lineshape of computation qubits leads to inefficient entanglement swapping.
3. The entanglement distribution rate between two QPUs in a single channel is fundamentally limited to the submegahertz regime due to the temporal overlap between adjacent single-photon wavepackets [17–19].
4. Existing optical switches suffer from tradeoffs among loss, crosstalk, speed, dimensionality, and bandwidth, which limit scalable QPU interconnectivity in a quantum data center [20–24].

The first challenge comes from the fact that most computation qubit platforms do not emit entangled photons in the low-loss telecommunication band. Since visible and near-infrared photons experience significantly higher loss in standard optical fibers, quantum frequency conversion (QFC) is widely used to convert photons from visible or near-infrared wavelengths into the telecommunication regime. This coherent process preserves quantum information and enables long-distance entanglement between heterogeneous computation qubits.

However, most existing QFC protocols focus exclusively on wavelength conversion, with limited attention to temporal mode shaping [25–27], leading to the second challenge mentioned above. The efficiency of photon–matter interactions is critically dependent on the temporal mode of the photonic qubit and the spectral lineshape of the computation qubit. In heterogeneous quantum networks, these spectral and temporal profiles are often mismatched, which degrades the entanglement swapping rate. Temporal shaping techniques such as quantum time lenses [25–27] or modification of emitter line shapes via the Purcell effect can potentially mitigate this mismatch, but remain challenging [28,29].

Although QFC has been successfully demonstrated, most implementations focus on one-to-one wavelength conversion and do not support multiplexing, resulting in a limited entanglement distribution rate due to the third challenge: the entanglement distribution rate per channel remains limited to the sub-megahertz regime between two QPUs. This constraint arises from the broad temporal wavepackets of entangled photons emitted by computation qubits [17–19]. To avoid inter-symbol interference between adjacent photons, the photonic qubit rate remains low, which directly limits the Bell pair generation rate. This is insufficient for executing error-corrected distributed quantum circuits, where high-fidelity Bell pairs need to be distributed at rates comparable to 1/10 of the $T_2$ time of computation qubits [30–32]. A quantum interface that supports spatial or wavelength multiplexing between computation qubits is therefore essential, and early work has begun to explore such approaches within limited bandwidths [33].

Unfortunately, the fourth challenge from optical switches can hardly support spatial multiplexing and also limits the execution of remote circuits. The scalability of quantum data centers is currently limited by the lack of optical switches that simultaneously offer low loss, low crosstalk, high speed, large bandwidth, and high port count (dimensionality). This limitation makes spatial multiplexing impractical and restricts the number of QPUs that can be interconnected at a single node. Mechanical switches provide low insertion loss and crosstalk across a broad bandwidth and high dimensionality, but their slow reconfiguration speed introduces latency that delays real-time quantum circuit execution [20–22,34,35]. Photonic switches, by contrast, can operate with picosecond to nanosecond reconfiguration times. However, their loss and crosstalk grow significantly with the number of switching elements [23,24,36]. Additionally, photonic switches typically operate over narrow spectral windows, making them incompatible with DWDM-based multiplexing.

Here, we propose a reconfigurable quantum interface (RQI) architecture that addresses the key limitations discussed above. The RQI enables on-demand tuning of both frequency and temporal modes of photonic qubits into specific DWDM channels with wavelength-selective switches (WSS), making it compatible with existing telecommunication infrastructure. This allows a high-rate entanglement distribution across heterogeneous QPUs, with theoretical rates up to 183.4 MHz and practical rates of 4.5 MHz under state-of-the-art experimental conditions.

To better illustrate our architecture and modeling of the networked computing systems, the rest of the paper is organized as follows. Section 2 introduces the architecture of using RQI and WSS to enable DWDM multiplexing at a low latency for distributed quantum computing. Section 3 focuses on the design and modeling of RQI devices based on $\chi^{(2)}$ and $\chi^{(3)}$ nonlinear converters. Section 4 shows our modeling of the WSS. Section 5 illustrates our modeling of the networked $^{87}$Rb quantum computing systems, and shows the simulation results of entanglement distribution rates and fidelity under different circuit execution scenarios. Section 6 concludes our work and extends the architecture to other major qubit modalities.

## 2. RQI ARCHITECTURE

Our architecture, illustrated in Fig. 1(b), is based on the emitter-emitter protocol in a quantum data center, and can be extended to other protocols as well [3]. Within each QPU, a subset of computation qubits is designated as communication qubits that generate matter–photon entangled pairs. Depending on the application, the architecture supports various photonic encodings, including Fock state, polarization, and time-bin. Following entanglement generation, the photonic qubit is routed through an array of reconfigurable quantum interfaces (RQIs), each comprising a nonlinear medium, tunable lasers, and control electronics. In the first stage, a narrow-linewidth, mode-hop-free tunable laser is adjusted to the appropriate phase-matching condition for efficient frequency conversion. A pulsed pump laser with a tailored spectrum is then used to reshape the temporal mode of the photonic qubit. By jointly tuning the pump wavelength, spectrum, and control parameters, the system can convert a photon from the native computation-qubit frequency and lineshape to an arbitrary DWDM channel frequency with the desired temporal profile. It is possible to achieve single-stage frequency and temporal mode conversion using a single nonlinear medium and a properly engineered pump [37]. After conversion, photons are multiplexed via a DWDM multiplexer and routed by a WSS to destinations, where they undergo Bell swapping with pho-tons from other QPUs to establish long-distance entanglement. Two configurations for switch networks are discussed in detail in Section 4.

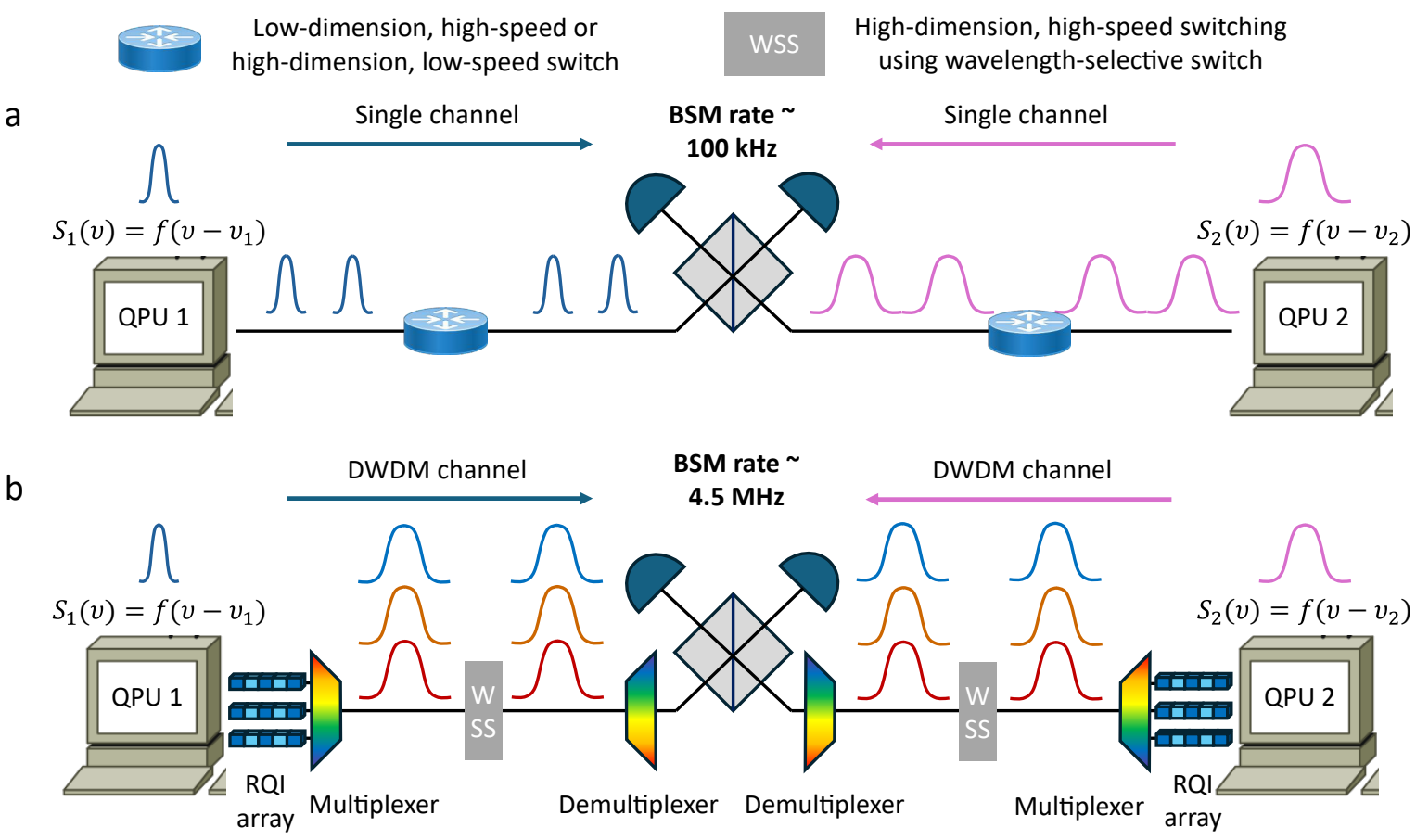

**Fig. 1.** (a) The conceptual illustration shows the current quantum network solution for heterogeneous entanglement swapping. The four existing challenges on different components of the quantum network are shown correspondingly. (b) The conceptual illustration of our scalable, programmable solution for distributing entanglement between heterogeneous quantum information processors, which includes quantum computers, memories, and sensors. WSS stands for wavelength-selective switch.

The programmability of the nonlinear interaction, based on either $\chi^{(2)}$ or $\chi^{(3)}$ processes, is achieved by tuning the refractive index of the medium and the spectral properties of the pump laser. The tuning mechanisms include thermal, electrical, optical, and optomechanical actuation, with response times ranging from nanoseconds to milliseconds [38–57]. Integrated, tunable pump lasers can be implemented via heterogeneous or hybrid integration of gain media, with on-chip control achieved through photonic integrated circuits [58,59]. These techniques enable the RQI to be reconfigured dynamically on an integrated photonic platform.

## 3. RQI MODELING

The noise of our approach comes mainly from RQI noise, DWDM multiplexer/demultiplexer crosstalk, and switch crosstalk. Most noises can be modeled from the results in the literature except for RQI noise [20–24,36,60–62]. Therefore, our noise and fidelity analysis mainly focuses on the simulation and possible improvement of RQI noise. In this section, we mainly focus on the frequency conversion part of the RQI, as the relevant research on quantum temporal mode conversion is limited. The diagram of the wavelength and noise spectrum in our discussion is shown in Fig. 2(a), including the most widely used single stage $\chi^{(2)}$ difference frequency generation (DFG) process, the low-noise $\chi^{(3)}$ converter based on a four-wave mixing Bragg grating (FWM-BG) process, and the theoretical noise-free $\chi^{(3)}$ converter based on the third order difference frequency generation (TDFG) process.

### A. RQI Based on $\chi^{(2)}$ Process

The most widely adopted $\chi^{(2)}$ material for QFC is the periodically poled Lithium Niobate (PPLN) waveguide, which has been demonstrated in single and two-stage conversions [14,15,63–82]. We use $^{87}$Rb-780 nm atom-photon entanglement as an example. $^{87}$Rb is a widely used platform in quantum computing and sensing [83,84], and its 780 nm transition can be efficiently coupled to optical cavities for high-fidelity photon collection. We thermally tune the conversion of the 780 nm photon to different DWDM channels, i.e., from 1519.86 to 1577.03 nm, following ITU standard 50 GHz grid. The corresponding pump laser wavelength range needs to be from 1543.33 to 1602.32 nm.

Due to the overlap of the frequency band of the converted idler photon with the frequency of the strong pump photon, this conversion is experimentally challenging using the $\chi^{(2)}$ process. The strong, broadband fluorescence and Raman scattering noise photon from the pump laser can significantly reduce the conversion fidelity. Compared to these two noise sources, noise photons from photoluminescence and spontaneous-parametric-down-conversion are usually much weaker and thus are ignored here. As shown in Fig. 2, Raman noise has a narrower spectral band than fluorescence noise but is usually stronger with asymmetric Stokes and anti-Stokes spectral noise peaks. The fluorescence noise is usually broader and stronger in the long wavelength regime of the pump, but the peak is generally weaker than the Raman noise. In addition to nonlinear noise, tuning noise, and mode impurity from the pump laser also contribute to the final infidelity of the converted photon. Another reported noise in the PPLN converter comes from the non-uniform poling period during the manufacturing process of the waveguide, which comes from the nanoscale inhomogeneity of the material and can hardly be controlled, especially for long waveguides.

Generally speaking, the major noise from the PPLN converter can be significantly reduced with a spectrally well-separated pump laser. Therefore, to minimize the noise level, we assume the use of thermal tuning within 20 DWDM channels from 1519.86 to 1527.22 nm, and the pump wavelength needs to be tuned from 1594.22 to 1602.32 nm. For such a wavelength range, the expected temperature tuning range is 27.26 degree Celsius with 0.27 nm per degree Celsius [64,85].

Even though the pump photon is about 80 nm away from the idler photon, as shown in Fig. 2(a), the fluorescence noise and Raman noise induced by the pump laser are sufficiently broadband to spectrally overlap with the idler photon. Since these spontaneous photons are spectrally indistinguishable from the coherent idler photon, the infidelity is fundamentally nonzero in this single-stage $\chi^{(2)}$ converter. The noise in the PPLN waveg-uide, designed for the conversion discussed above, is shown in Fig. 2(b). Under strong pumping conditions, the internal quantum conversion efficiency can easily reach unit efficiency, but the noise flux is greater than $10^6$ photons per second per nm. Obviously, a longer pump wavelength introduces significantly lower nonlinear noise counts. However, in the $\chi^{(2)}$ process, such flexibility is not available. Furthermore, the tuning noise can add on top of the nonlinear noise as well, leading to an overall noise above $10^7$ photons per second per nm as shown in the subfigure of Fig. 2(b). Without filtering, the noise photon can easily overwhelm the converted idler photon in PPLN converters. Unfortunately, even with strong filtering, as has been demonstrated [14,15,78], the noise photon can hardly be reduced to below 50 counts per second.

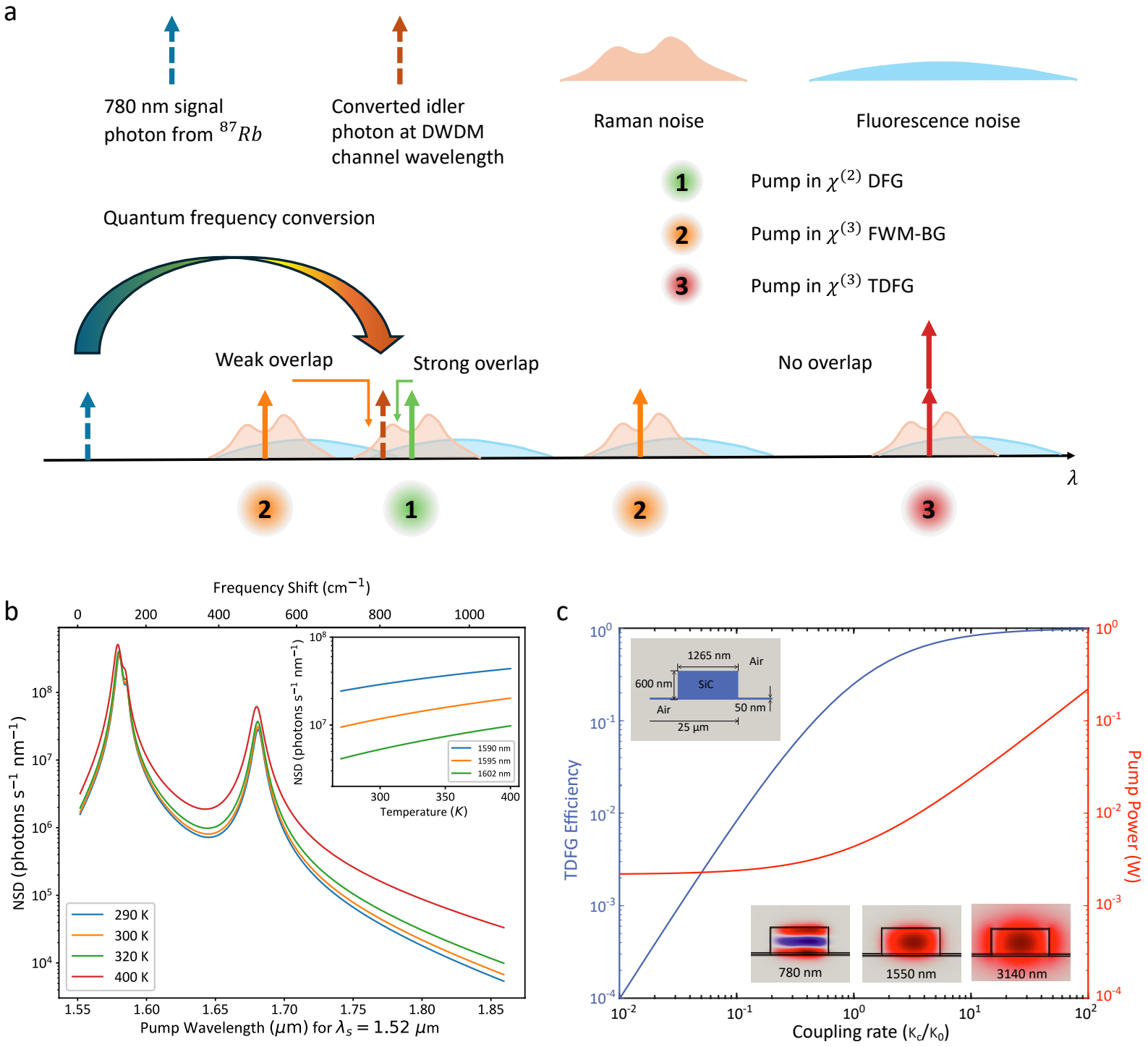

**Fig. 2.** (a) Quantum frequency conversion diagram for three nonlinear processes: $\chi^{(2)}$ DFG, $\chi^{(3)}$ FWM-BG, and $\chi^{(3)}$ TDFG. The major noises analyzed here are pump-induced Raman noise and fluorescence noise. A spectral overlap between the noise and the dashed arrow indicates that the noise can be found in the conversion. (b) Noise in a $\chi^{(2)}$ converter as a function of pump wavelength and tuning temperature. (c) Conversion efficiency and required pump power in a $\chi^{(3)}$ TDFG converter. The device (the top inset) is a silicon carbide microring with a thin bottom support layer with air top and bottom cladding. The optical modes used (the bottom insets) are third-order-vertically transverse-electric mode at 780 nm, and fundamental transverse-electric modes at 1550 and 3140 nm.

### B. RQI Based on $\chi^{(3)}$ Process

Our example shown in the previous section illustrates that, under some configurations, the fidelity of converted idler photons in the $\chi^{(2)}$ process is fundamentally compromised due to the spectrally indistinguishable spontaneous photons. Although some promising approaches are proposed to reduce the noise photon level [14,15,78,86], significantly higher overall loss is introduced as well, which brings the $\chi^{(3)}$ based converter into our consideration.

The $\chi^{(3)}$ process has been investigated for frequency conversion in the past decade, and a possible solution is to choose the spectral location of the two pump laser frequencies in the FWM-BG process to reduce the noise, shown in Fig. 2(a), which has been experimentally demonstrated in both fiber and photonic resonators with over 60% efficiency [87–89]. As shown in Fig. 2, the nonlinear noise from spectrally well-separated pump lasers is significantly lower than the nonlinear noise in the $\chi^{(2)}$ converter. However, the fluorescence noise induced by the pump with a wavelength between that of the signal and the idler photon is usually very broad in the long-wavelength regime, and this noise photon can leak into the converted idler photon bandwidth (shown as the leftmost blue noise in Fig. 2(a)). Therefore, this approach still yields, although generally orders of magnitude lower than that of the $\chi^{(2)}$ process, some amount of noise, resulting in a limited fidelity.

Alternatively, a noise-free approach based on a third-order difference/sum frequency generation process has been theoretically proposed [90]. By using two degenerate photons from a mid-IR laser, one can convert the 780 nm photon to the telecommunication band with minimal impact from noise photons generated by high-power lasers, including, but not limited to, Raman scattering, spontaneous FWM noise, and fluorescence noise. Figure 2(c) shows the simulated photon number efficiency of converting a 780 nm photon to a 1550 nm photon using a pump laser at 3140 nm, and the required pump power assuming that the pump laser is critically coupled. To achieve a 90% conversion efficiency, the microring needs to be 20 times overcoupled at 780 and 1550 nm, and the required pump power is only 46 mW. Here, we use high-quality 4H-crystalline silicon carbide as the core material [91,92] with air claddings on top and bottom. This configuration is used not only to confine the optical modes in silicon carbide but also to avoid the noise generation from amorphous materials such as silicon nitride and silicon ox-ide [93]. While this simulation result is accurate in concept, we note that the realization of the coupling conditions at these three wavelengths, while maintaining high optical intrinsic qualities, as well as dispersion matching of three modes, requires non-trivial engineering efforts in design and fabrication. In order to do this, it is practical to use thermal tuning of the ring resonator to relax the dispersion tolerance. For example, using integrated heaters [94], the dispersion matching of all three wavelengths can be tuned by a few tens of nanometers, which is crucial for practical realizations of efficient RQI at targeted wavelengths.

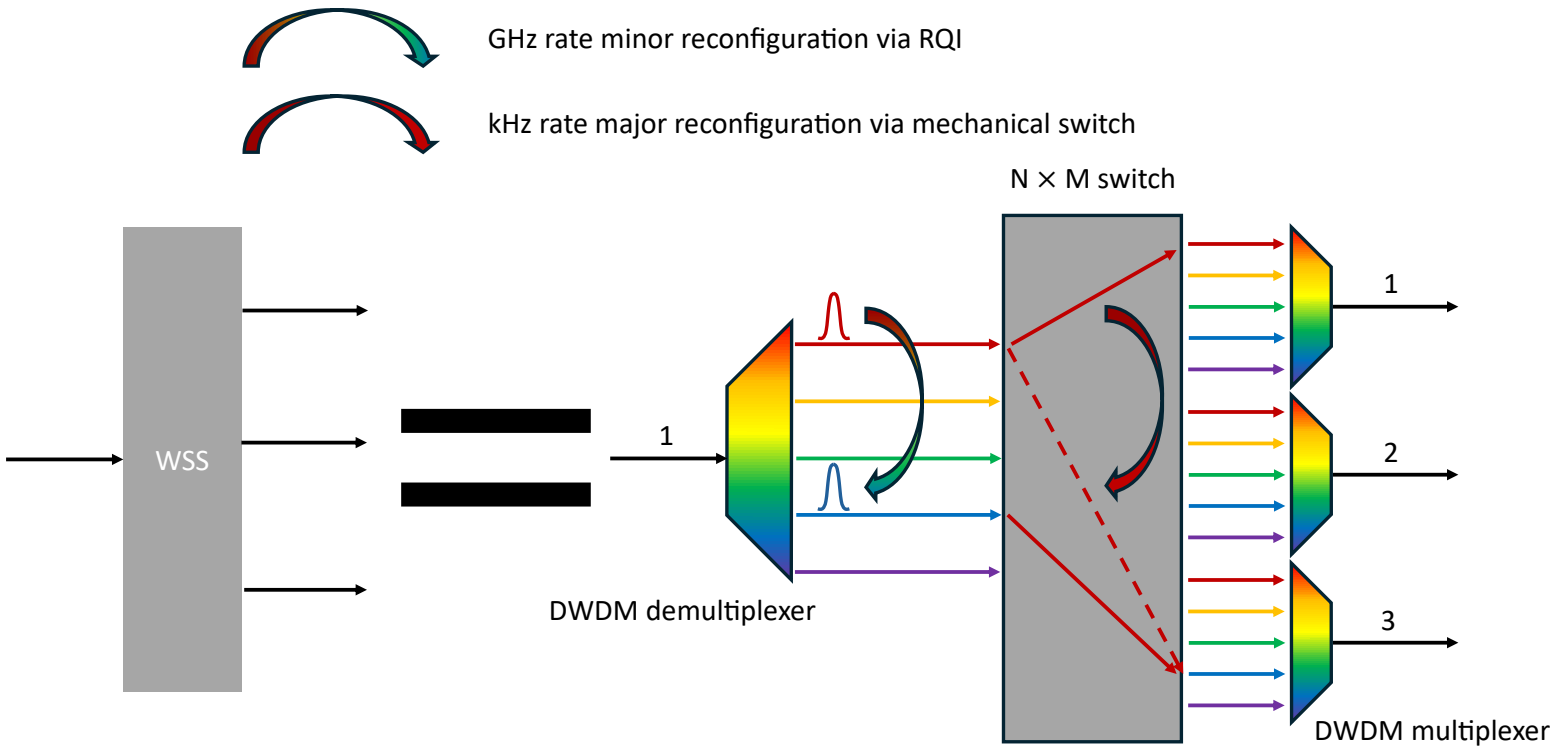

**Fig. 3.** The illustration of two types of reconfiguration using the RQI array and a wavelength-selective switch (WSS). Each photon passes through two DWDM multiplexer/demultiplexers and one $N \times M$ switch. The $N \times M$ switch provides all-to-all connectivity, and the $M$ output ports are connected to different DWDM multiplexers. In this example, the $N \times M$ mechanical switch connects ports via solid red lines, and the user needs to reroute the photon from output port 1 to output port 3. If the minor configuration is used, the reconfiguration can be rapidly finished by tuning the photon from the red DWDM channel to the blue DWDM channel at GHz level. If the major reconfiguration is used, the $N \times M$ mechanical switch reconfigures the connection from the solid red line to the dashed red line at kHz rate.

## 4. IMPROVED NETWORK RECONFIGURABILITY WITH WSS

Instead of using standard mechanical or photonic switches, wavelength-selective switches (WSSs) are introduced to significantly improve network reconfigurability. Although previous studies have mentioned the potential to improve network management via WSS [7,8,11,95–98], our architecture is not limited to broadband entanglement sources, and can be applied to various modality of entanglement source.

In our architecture, we model a WSS as shown in Fig. 3, which consists of DWDM multiplexers/demultiplexers and one $N \times M$ switch. The reconfiguration comes from two aspects, leveraging the WSS and the fast reconfiguration speed of RQI. The first aspect of reconfiguration, which is the major reconfiguration, comes from the $N \times M$ switch inside the WSS, which can offer high-dimension, low-loss, low-crosstalk, and non-blocking all-to-all connectivity using a conventional mechanical switch [21,22]. The second aspect of reconfiguration, called minor reconfiguration, comes from programming the RQI, whose tuning speed can be up to ns [40,45]. In a minor reconfiguration, the $N \times M$ mechanical switch remains static, but the DWDM channel assigned to each RQI module is changed. This reconfiguration offers limited connectivity but much faster speed, which can satisfy the 2-qubit gate execution time. In the next section, we will explicitly show that, with this modeling of the WSS, the low latency enables an average entanglement distribution successful time comparable to or less than 1/10 of the $T_2$ time of computation qubits [30–32]. In a major reconfiguration, both the mechanical switch and the assignment of the DWDM channel are reconfigured, providing arbitrary connectivity, but the rate is limited by the speed of the mechanical-switch reconfiguration, which is usually at the kilohertz level.

In distributed quantum computing (DQC), one main goal in circuit partition is to reduce the distribution of long distance entanglement by carefully partitioning the circuit and scheduling the job [99]. In such a scenario, one can ensure that the re-assignment of entanglement between QPUs happens mostly inside each rack, or between neighboring racks [35]. Therefore, during the execution of each job, the reconfiguration of the switch needs to be rapid enough to accommodate 2-qubit gate execution time, but not dramatic, and most connections in the network are not reconfigured. Hence, a minor reconfiguration is usually sufficient during DQC job execution. When loading the new job, based on the partition result, the entire network needs to be reconfigured dramatically, which usually requires a high connectivity in the network, and thus a major reconfiguration is necessary. Since this only happens between DQC jobs, a slow but high-connectivity major reconfiguration does not lead to a significant delay for circuit execution. Considering that most useful circuits require lots of distributed Bell pairs, the desired major reconfiguration only happens every couple of ms, which is not challenging for mechanical switches [100].

Loss plays a significant role in quantum networking. Based on previous experimental results, this WSS model can suffer a loss of less than 1.5 dB, and a theoretical design has shown the potential to reduce the loss to less than 0.7 dB with a specifically designed DWDM multiplexer/demultiplexer [62]. Compared to the photonic-switch-only approaches, the loss and crosstalk of our approach is significantly lower, while the minor reconfiguration speed is at the same level. This result is reached mainly because increasing the dimension of a mechanical switch does not significantly increase the loss. However, increasing the dimension in a photonic switch usually leads to more loss as the PIC depth goes deeper. Compared to mechanical-switch-only schemes, our approach sacrifices some loss and intra-job network connectivity to gain a significantly faster reconfiguration speed, leading to a much less job execution time for distributed quantum circuits.

In our architecture, when a new job is going to be loaded, one first optimizes the partition of the circuit to reduce the number of cross-rack and cross-data center entanglement distributions. Based on this result, one performs a major reconfiguration of each WSS in the network to best accommodate the entanglement distribution for this DQC job. During the job, one only reconfigures the network by programming the RQI in the corresponding QPUs.

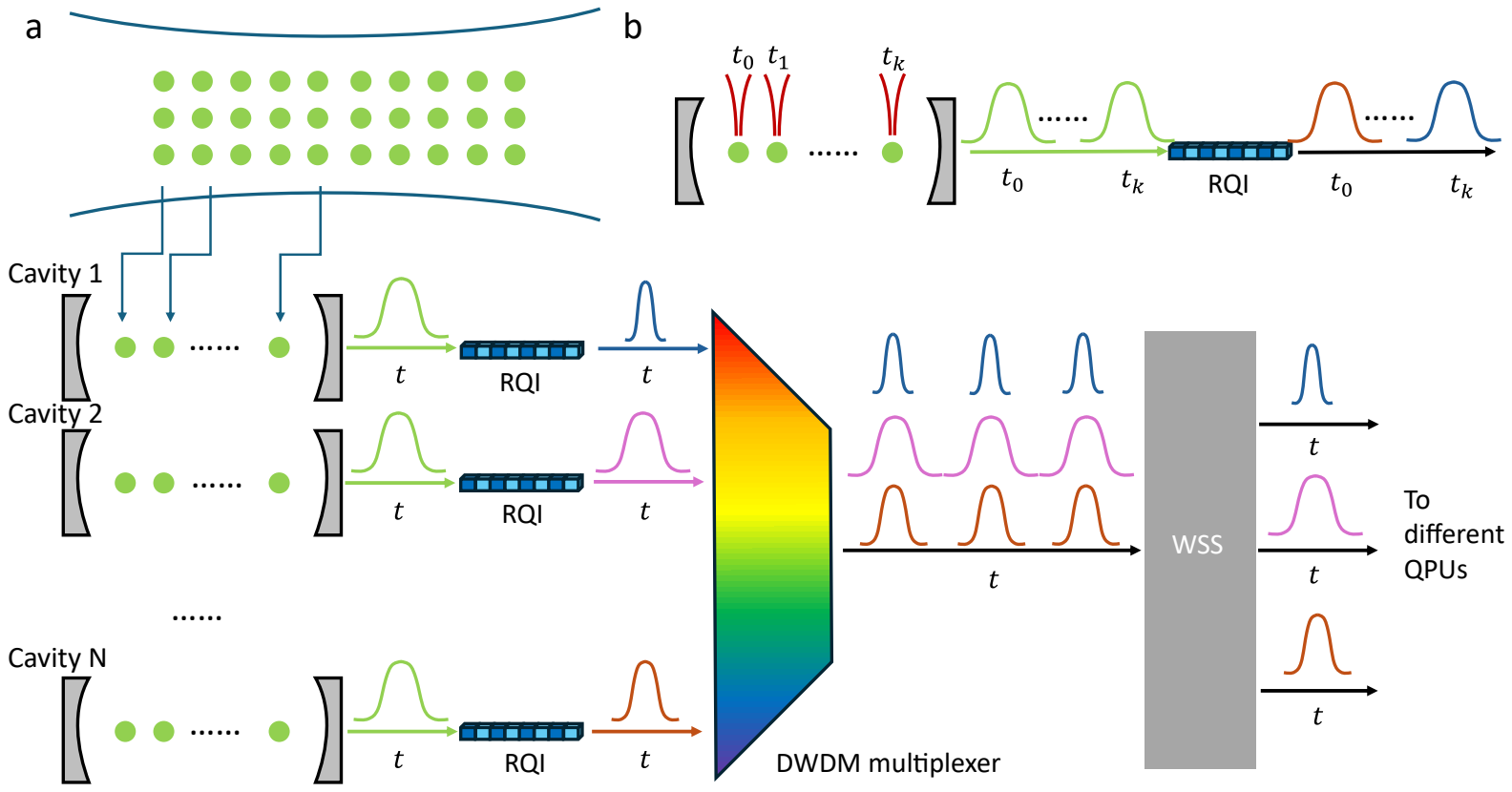

**Fig. 4.** (a) The schematic of the RQI system based on $^{87}$Rb atoms. Atoms are loaded into microcavities via shuttling. The frequency and temporal mode of entangled photons after each microcavity are programmed to the desired frequency and temporal mode to match the photon used in the Bell swapping procedure. All entangled photons are MUXed via a multiplexer to fully leverage the rate advantage of the standard DWDM protocol. The WSS is used to improve the network reconfigurability, which is discussed in more detail in Section. 4. (b) The top right figure illustrates the generation of an atom-photon entanglement pair using the TDM approach. In each microcavity, each atom is prepared and excited sequentially, resulting in the optimal single-channel atom-photon entanglement rate. The design and realization of RQI can be found in Section. 3 and Fig. 2.

## 5. ENTANGLEMENT DISTRIBUTION RATES AND FIDELITY BETWEEN NETWORKED QPUs

We evaluate the entanglement distribution performance of our RQI-based architecture using $^{87}$Rb atoms as communication qubits in a cold-atom quantum computing system. Our network modeling is based on the modeling of RQI and WSS in the previous sections, and the cold atom computer modeling is based on the architecture described in Ref. [101].

As illustrated in Fig. 4(a), the $^{87}$Rb atoms are divided into $N$ groups, each corresponding to a DWDM channel. Each group contains $k$ atoms, giving a total number of communication qubits $N_{\text{tot}} = kN$. Group assignments are determined by circuit execution and partition requirements. Each group is loaded into a high-finesse optical microcavity with efficient photon collection, and each cavity is associated with a unique DWDM frequency. The number of groups matches the number of available DWDM channels. Microcavity arrays for scalable cold-atom quantum systems have been demonstrated experimentally [102–107].

Once atoms are loaded into their cavities, they are sequentially excited following a time-division multiplexing (TDM) protocol [17–19,108,109], as shown in Fig. 4(b). Entangled photons emitted at 780 nm are collected from the cavity, coupled into single-mode fibers, and routed to integrated RQI modules for frequency and temporal mode conversion.

We model the entanglement distribution under three key scenarios in a quantum data center: intra-rack, neighboring-rack, and cross-data-center. The entanglement distribution rates and fidelity for each scenario are simulated with three network configurations: (i) single channel without quantum frequency conversion (QFC), (ii) single channel with QFC, and (iii) RQI with DWDM multiplexing. The temporal mode converter is ignored here since all computation qubits are $^{87}$Rb.

In an idealized system, the entanglement distribution rate between two cavities is bounded by 229.3 kHz due to the TDM protocol [17,18], giving a theoretical maximum of 183.4 MHz across all DWDM channels. In practice, optical component losses, finite switch latency, and limited DWDM filter efficiency reduce the achievable rate. With realistic parameters, our modeling shows a practical entanglement distribution rate of up to 4.5 MHz, The simulation is based on an event-by-event Monte Carlo simulation pipeline, and the details can be found in Supplement 1, Section 4.

In intra-rack configurations, a small number of QPUs, typically a few dozen, can execute low-depth circuits with photonic links spanning only a few meters. A single top-of-rack switch provides full connectivity, but must operate on microsecond or even smaller timescales to match two-qubit gate execution speeds [30–32]. As shown by the red curves in Fig. 5(b), both single-channel configurations, with or without QFC, saturate at an entanglement distribution rate of approximately 25.2 kHz, which falls short of the throughput required for error-corrected circuit execution. This saturation occurs when the number of communication qubits exceeds 100. In contrast, our RQI-DWDM architecture achieves more than 4.5 MHz, maintaining near-linear scaling with $N_{\text{tot}}$ up to 1000 communication qubits. This gain arises from parallel DWDM channel multiplexing and nanosecond-scale reconfigurability.

For larger-scale circuits, multiple QPU racks are required. Previous studies show that most benchmark quantum workloads fall into this category [35]. To minimize fiber length and switch overhead, distributing entanglement between neighboring racks, typically separated by tens of meters, is preferred. Although circuit partitioning can reduce inter-rack communication, the limited qubit capacity per QPU makes it difficult to localize all Bell pair generation [35,99]. This topology requires multiple switches to support all-to-all inter-rack connectivity. As shown by the yellow curves in Fig. 5(b), the single-channel QFC con-figuration achieves only 13.3 kHz, and without QFC, the rate drops to 9.7 kHz, both too slow to meet real-time execution requirements. In contrast, our RQI-DWDM architecture delivers up to 3.8 MHz, more than two orders of magnitude faster. These gains again result from efficient wavelength multiplexing and fast, low-loss switching. For most distributed quantum applications, this ensures that the network meets latency constraints for most distributed quantum computing jobs.

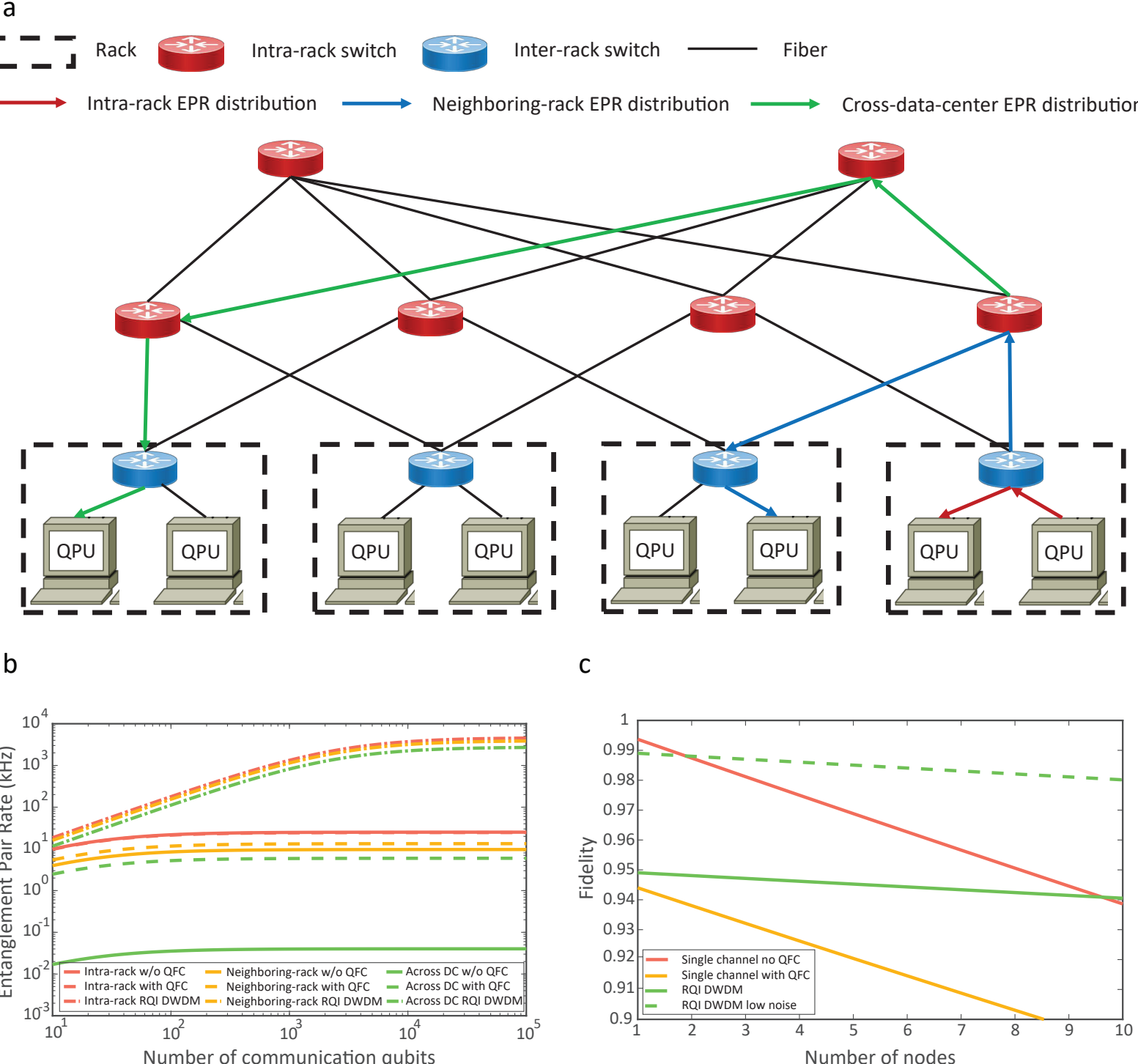

**Fig. 5.** (a) An example of quantum data center based on Clos network topology. Three different distributed quantum computing scenarios are illustrated: intra-rack, neighboring-rack, and cross-data-center. (b) Entanglement distribution rate for single channel without QFC, single channel QFC, and RQI DWDM. (c) Fidelity of distributed photonic qubits as a function of the number of nodes in the network. Entangled photons are assumed to be emitted at an optimal rate from each QPU. Details of the simulation pipeline can be found in Supplement 1, Section 4.

In large-scale circuits requiring millions of physical qubits, entanglement must be distributed across data centers. In this regime, the fiber lengths extend to several kilometers, introducing significant attenuation and latency. Without QFC, the entanglement distribution rate drops to just 40 Hz, as shown by the green curves in Fig. 5(b), making the network the dominant bottleneck. The single-channel QFC improves the rate modestly to 5.9 kHz, but this remains much below the threshold needed for the timely execution of distributed algorithms. In contrast, our RQI-DWDM architecture sustains a rate of 2.7 MHz under the same conditions, more than three orders of magnitude higher, sufficient to support both gate execution and error correction. These results demonstrate that even in cross-data center deployments, our architecture preserves the feasibility of distributed quantum computing without network-induced delays.

We model the impact of network-induced noise on entanglement fidelity, including contributions from RQI modules, switch crosstalk, and DWDM components, as shown in Fig. 5(c). In small-scale networks, where photons traverse only a few nodes, our architecture initially exhibits higher noise due to the use of a $\chi^{(2)}$-based RQI, as modeled in Section 3.1, with non-negligible conversion noise.

However, as the network scales and more nodes are involved, our architecture becomes increasingly robust compared to single-channel configurations. This is mainly because high-speed photonic switches, required to maintain low-latency circuit execution, introduce significantly more crosstalk than mechanical alternatives [100]. In conventional designs, this crosstalk accumulates with each switch traversal, degrading fidelity. In contrast, our architecture's multiplexing reduces the number of required switch layers, limiting total noise accumulation.

As discussed in Section 3.2, the noise introduced by RQI can be significantly reduced through careful engineering of the nonlinear interaction. With an improved RQI design, our approach can surpass single-channel fidelity even in small networks, as illustrated by the green dashed curve in Fig. 5(c). Overall, our architecture provides high-rate and high-fidelity entanglement distribution, especially as network complexity grows.

## 6. DISCUSSION AND CONCLUSION

Although our example is based on cold atoms, the architecture naturally works for other major quantum computing platforms.

Trapped ions constitute a major platform for quantum computing, featuring the highest gate fidelity and all-to-all connectivity. However, scaling up the size of a trapped-ion quantum computer is very challenging, and connecting several QPUs within a quantum network is considered a promising solution. Similarly, to neutral atoms, trapped-ion qubits can hardly be optically accessed using photons in the telecommunication bands, and thus RQI methods can be used in a useful way [67,69,73,75,76,80].

**Table 1. Comparison of Distributed Quantum Computing (DQC) Metrics between Our RQI Architecture and Current Solutions**

| DQC Metrics | Single Channel w/o QFC | Single Channel with QFC | RQI DWDM |
|---|---|---|---|
| Intra-rack Rate (kHz)[a] | 25.16 | 24.66 | 4508[b] |
| Inter-rack Rate (kHz) | 9.66 | 13.34 | 3844 |
| Cross-DC Rate (kHz) | 0.040 | 5.96 | 2696 |
| Fidelity[c] (3 nodes) | 0.987 | 0.938 | 0.946 |
| Fidelity[c] (9 nodes) | 0.924 | 0.878 | 0.926 |
| Latency | ~ns | ~ns | ~ns |
| Connectivity | All-to-all | All-to-all | All-to-all[d] |

[a] Assume use GHz reconfiguration rate photonic switch.
[b] Theoretical upper bound is at 183.4 MHz.
[c] Assume the atom-photon entanglement fidelity is equal to 1.
[d] All-to-all connectivity in major reconfiguration.

The shuttling process in our example can be simplified by assigning each ion chain to an RQI module, and the rest of the system is similar to what we proposed in Fig. 4.

Our architecture is naturally compatible with color-center quantum memories, including SiV, NV, and SnV. One major challenge of color centers comes from the variation in resonance frequencies for different color centers. Although recent demonstrations have shown possible tuning of the resonance, an average of the 2 GHz tuning range can hardly mitigate the resonance mismatch between different color centers [110]. Considering that QFC has been demonstrated to be essential for color centers in quantum networks [26,77,81,82], our approach can solve frequency variations in different color-center samples, as well as significantly improve the entanglement rate. In terms of implementation, the RQI array can be integrated with the color-center nanophotonic cavities with appropriate couplers, i.e., replace the microcavity array in Fig. 4 with nanophotonic cavities.

Quantum dots are considered to be a promising candidate for deterministic single-photon emission for many quantum information applications. However, fabricating identical quantum dots remains a challenge, and the wavelength is usually randomly distributed around the desired wavelength because of spectral diffusion. To interfere photons from multiple quantum dots, QFC needs to be implemented. Therefore, with RQI, one can multiplex and Bell swap photons emitted by spectrally distinguishable dots without concern of fabrication defects.

In summary, we propose a quantum networking architecture based on RQI and WSS to improve the entanglement distribution for distributed quantum computing. As summarized in Table 1, compared to current quantum network architectures, our architecture can improve the entanglement distribution rate to 183.4 MHz, using the industrial standard DWDM protocol. By dropping inessential connectivity within the circuit execution, the reconfiguration time can be reduced down to the nanosecond level without sacrificing the entanglement distribution rate and fidelity. In terms of the RQI design and modeling, we analyze the infidelity induced by noisy $\chi^{(2)}$ converters and propose a theoretically noise-free $\chi^{(3)}$ converter design. With consideration of entanglement distribution rate and fidelity, network management, circuit partition, and execution, the modeling results show that our architecture can satisfy the requirement for distributed quantum computing and error correction. Our approach is based on the standard DWDM infrastructure, which indicates a promising low-cost solution to enable distributed quantum computing.

Note that during our draft preparation, we found a proof-of-concept demonstration from Osaka University [111].

**Acknowledgment.** The authors acknowledge the helpful discussion from Dr. Hassan Shapourian, Prof. Martin Fejer at Stanford University, and Prof. Galan Moody at the University of California, Santa Barbara, as well as the support of PPLN waveguide design and fabrication from Dr. Shijie Liu and Prof. Yuanlin Zheng at Shanghai Jiao Tong University. Jiapeng Zhao, Eneet Kaur, Michael Kilzer, and Reza Nejabati are listed as innovators of the US patent application (serial number: 18/952,548) on the reconfigurable quantum interface.

**Disclosures.** Jiapeng Zhao, Eneet Kaur, Michael Kilzer, and Reza Nejabati are listed as innovators of the US patent application (serial number: 18/952,548) on the reconfigurable quantum interface.

**Data availability.** All data generated and analyzed during this study are available from the corresponding author on reasonable request.

**Supplemental document.** See Supplement 1 for supporting content.